# Study of the Impact of the Big Data Era on Accounting and Auditing

Yuxiang Sun[1], Jingyi Li[2], Mengdie Lu[3], Zongying Guo[4]

[1]EY, 100 Grand Ln, Foster City, CA 94404, USA
[2]BDO USA P.C., Los Angeles, CA 90071, USA
[3]Westcliff University, Irvine, CA 92614, USA
[4]The Hong Kong Polytechnic University, Fremont, CA 94539, USA

**Abstract:** In the wake of rapid advancements in science and technology, big data has emerged as a pivotal facet of contemporary society, revolutionizing the way we collect, analyze, and utilize information. Its integration into accounting and auditing practices has witnessed a steady and expansive trajectory, fundamentally reshaping traditional approaches to financial analysis, risk assessment, and compliance monitoring. This paper aims to delve into the profound influence of the big data era on the realms of accounting and auditing, meticulously dissecting the challenges and opportunities that accompany this evolution. As organizations grapple with unprecedented volumes of data generated from diverse sources such as social media, IoT devices, and transactional records, the traditional methods of financial reporting and auditing are being challenged. The sheer volume, velocity, and variety of data require innovative approaches to extract meaningful insights while ensuring data integrity and reliability. Moreover, the emergence of artificial intelligence and machine learning algorithms offers promising avenues for automating routine tasks, enhancing analytical capabilities, and detecting anomalies with greater accuracy and efficiency. However, the proliferation of big data also presents formidable challenges, ranging from data privacy and security concerns to the complexities of data integration and interoperability. Accounting and auditing professionals must navigate this complex landscape while upholding ethical standards, regulatory compliance, and professional skepticism. Moreover, the rapid pace of technological change necessitates continuous learning and adaptation to stay abreast of emerging trends and best practices. In light of these challenges and opportunities, this paper endeavors to proffer strategic countermeasures tailored to navigate this transformative landscape effectively. From leveraging advanced analytics tools and cloud computing platforms to investing in cybersecurity measures and talent development initiatives, organizations can enhance their capacity to harness the potential of big data while mitigating associated risks. Moreover, collaboration between stakeholders, including regulators, industry associations, and academia, is essential to foster innovation, knowledge sharing, and industry-wide standards. By embracing the opportunities afforded by the big data era while addressing its inherent challenges, accounting and auditing professionals can position themselves as trusted advisors, equipped to provide actionable insights and strategic guidance in an increasingly data-driven world. Through continuous collaboration, innovation, and adaptation, the profession can seize the transformative potential of big data to drive sustainable growth, enhance transparency, and uphold the integrity of financial markets.

## 1. Introduction

In the surge of the big data era, we witness a relentless push towards unprecedented heights in data acquisition, processing, and analysis. This tidal wave of information catalyzes profound transformations within the realms of accounting and auditing. Traditional methodologies are being upended as big data technologies permeate every aspect of our work, delivering a quantum leap in efficiency and accuracy. The advent of big data analytics has enabled accountants and auditors to extract insights from vast troves of data that were previously unimaginable. By harnessing sophisticated algorithms and machine learning techniques, professionals can now identify patterns, trends, and anomalies in financial data with unparalleled speed and precision. This newfound capability not only streamlines auditing procedures but also enhances the quality of financial reporting, providing stakeholders with a more comprehensive understanding of a company's financial health. Furthermore, the integration of big data into accounting and auditing practices facilitates real-time monitoring and risk assessment. With access to a continuous stream of data from various sources, auditors can promptly identify emerging risks and adjust their audit procedures accordingly. This proactive approach not only enhances the effectiveness of audits but also enables companies to mitigate potential risks before they escalate into crises. However, amidst the vast potential of big data, formidable challenges emerge. The specters of data security breaches and privacy infringements loom large, serving as constant reminders of the dual-edged nature of this data-driven revolution. As auditors handle increasingly sensitive and confidential information, the risk of unauthorized access or data manipulation becomes a significant concern. Moreover, the proliferation of regulations such as the General Data Protection Regulation (GDPR) and the California Consumer Privacy Act (CCPA) adds another layer of complexity, requiring auditors to navigate a labyrinth of legal requirements to ensure compliance. Against this backdrop, this paper delves into the profound impacts of the big data era on accounting and auditing practices. We scrutinize how big data technologies reshape workflows, enhancing not only efficiency but also precision. Additionally, we confront the critical issues of data security and privacy protection that accompany this era, aiming to provoke thoughtful consideration and strategic foresight for the future evolution of accounting audit practices. Through a comprehensive analysis of both the opportunities and challenges presented by the big data revolution, we seek to provide insights that will empower accounting and auditing professionals to navigate this transformative landscape with confidence and foresight.

## 2. The Impact of the Era of Big Data on Accounting

### 2.1. The changes in accounting data processing

In the wave of the big data era, the sourcing of accounting data has undergone seismic shifts compared to traditional practices. In the conventional model, accountants relied solely on data generated within the enterprise to compile and scrutinize financial statements. Internal enterprise data formed the cornerstone of accounting efforts, holding a pivotal position encompassing daily financial transactions, inventory flux, employee compensation, fixed asset depreciation, and more [1]. These data functioned as the enterprise's vital signs dashboard, offering a clear glimpse into its operational, financial, and cash flow statuses. However, amidst the deluge of big data, external market data has surged forth to become an integral component of accounting endeavors. External data, including industry trends, competitor analyses, market price fluctuations, and consumer behavior trends, act as market weather vanes, furnishing enterprises with invaluable market intelligence and competitive insights. By delving into these datasets, enterprises gain a sharper understanding of market opportunities, enabling agile adjustments to business strategies and facilitating more informed decision-making among key stakeholders. In this era of big data, the demands placed on accounting data processing grow increasingly rigorous. Accountants must swiftly respond to and process the latest data in real-time, ensuring that enterprises remain agile in rapidly evolving market landscapes. This dynamic necessitates accountants to closely monitor data fluctuations, continually refine and optimize analytical methodologies and conclusions, and uphold the timeliness and precision of accounting operations.

### 2.2. The challenges of accounting standards and systems

The advent of the big data era has unleashed unprecedented impacts and challenges upon accounting standards and systems. Traditional methodologies find themselves strained to their limits when confronted with the novel complexities of big data environments. Historically, accounting standards and systems have been anchored in conventional data processing and analysis techniques, adept at handling small-scale, well-defined datasets. However, the onset of big data has heralded an exponential surge in data volume and intricate data structures, replete with vast troves of unstructured data like texts, images, and videos. This deluge renders accountants incapable of faithfully depicting the true financial landscape and operational performance of enterprises using traditional frameworks. Moreover, the big data milieu has birthed a plethora of emergent issues—data ownership, usage rights, privacy protection, to name a few—that remain inadequately addressed within the confines of traditional accounting norms. To remain relevant in the era of big data, traditional accounting standards and systems necessitate profound recalibration and enhancement. Thankfully, the rapid evolution of big data technology presents a silver lining, furnishing fresh opportunities and expansive vistas for the evolution and implementation of accounting standards and systems. Big data technology harnesses its formidable data processing and analysis prowess to distill valuable insights from vast datasets, furnishing more robust and holistic support for the formulation of accounting standards and systems. By mining and analyzing external data such as market trends and industry benchmarks, we can more accurately assess pivotal indicators like asset valuation and enterprise risk profiles, facilitating the crafting of more pragmatic and effective accounting frameworks. Furthermore, big data technology can be leveraged extensively in the enforcement and oversight of accounting standards and systems. Through its deployment, we can monitor and scrutinize enterprise financial statements in real-time, swiftly identifying and deterring potential financial malfeasance and irregularities. This proactive approach markedly enhances the efficiency and efficacy of accounting standards and systems enforcement, ensuring greater integrity and transparency in financial reporting.

### 2.3. Accounting information security and privacy protection

In the era dominated by the overwhelming influx of big data, the imperative for safeguarding accounting information's security and privacy has been thrust into the spotlight. Confronted with vast reservoirs of data, enterprises find themselves compelled not only to efficiently harness this trove to propel business advancement but also to meticulously safeguard data integrity and confidentiality, ensuring the fortification of enterprise assets and preservation of their hard-earned reputation. In this digital milieu, the scope of accounting data transcends conventional financial statements and ledgers, spanning a broad spectrum of domains including customer transaction records, supply chain intricacies, market analytics, and beyond, all brimming with immense commercial significance. As the value of data skyrockets, so do the perils it encounters, ranging from malevolent hacking attempts to clandestine internal breaches and unauthorized access. A single misstep can precipitate data breaches, entailing immeasurable losses for enterprises. Hence, it behooves enterprises to adopt pragmatic and efficacious measures to bolster the security mantle enveloping accounting information. Employing data encryption technology emerges as a pivotal strategy in this endeavor. By encrypting sensitive data, even in transit or storage, the content remains inscrutable to malevolent actors, thwarting their nefarious designs. Furthermore, access control technology emerges as a linchpin in ensuring data integrity. By delineating distinct access privileges for disparate users and rigorously constraining the handling and interaction with sensitive data, it serves as a bulwark against internal malfeasance and external encroachments. This proactive approach empowers enterprises to mitigate the risks inherent in an increasingly data-centric landscape, fortifying their resilience against evolving threats and safeguarding the sanctity of accounting information.

## 3. The Impact of the Era of Big Data on Auditing

### 3.1. The innovation of audit methods and technologies

The era of big data has injected unprecedented vitality and opportunities into the methodology and technological innovation in the audit field. In this age, auditors harness the powerful analytical capabilities of big data technology to delve deep into datasets, uncovering hidden risks and issues lurking beneath the surface. Unlike traditional auditing methods constrained by the limitations of data scale and processing speed, which often rely on sampling data, auditors now have the capacity to conduct comprehensive analyses across all available data. This shift towards an all-data audit approach significantly enhances the breadth and precision of audits, bringing to light subtle risks and problems that previously evaded detection. Moreover, big data technology facilitates the optimization of audit processes and the enhancement of efficiency. By automating tedious tasks such as data sorting, screening, and comparison, auditors can redirect their time and expertise towards matters demanding professional judgment and in-depth analysis. This transformation vastly improves audit efficiency, mitigates errors and oversights attributable to human factors, and elevates the overall quality and reliability of audit outcomes [2].

### 3.2. The changes in audit evidence and procedures

Audit work is evolving in response to the changing landscape of audit evidence. Traditionally, auditors relied on financial statements, vouchers, account books, and other paper documents from enterprises. However, the advent of big data has ushered in a new era, where electronic data plays a pivotal role alongside traditional sources. The efficiency, convenience, and ease of storage offered by electronic data make it an integral component of audit evidence. In today's environment, auditors must adapt to the influx of diverse data sources, including social media information, online transaction records, and Internet of Things data. These sources offer fresh perspectives and entry points for audit investigations. Adhering rigidly to old methodologies is no longer viable; auditors must expand their horizons and embrace big data technologies. This entails actively acquiring the skills to sift through vast datasets, extracting valuable insights to bolster audit evidence. To succeed in this new paradigm, auditors must possess not only a robust foundation of professional knowledge and practical experience but also sharp insight and an innovative mindset. Audit procedures must also evolve to suit the complexities of big data. Traditional approaches fall short in the face of massive datasets, necessitating adjustments and optimizations tailored to the big data environment.

## 4. The Coping Strategies and Suggestions

### 4.1. The strengthening of research and development and application of big data technolog

With the wave of digitalization sweeping the world, the significance and utilization of big data technology by enterprises have emerged as pivotal factors in driving industry development. Within the realm of accounting and auditing, the profound integration and innovative deployment of big data technology hold immeasurable promise for enhancing financial management efficiency, fortifying risk prevention and control capabilities, and elevating decision-making support across enterprises. Consequently, enterprises must proactively allocate resources toward research and development, thereby fostering the widespread adoption of big data technology within the accounting and auditing domains. However, this evolution cannot be solely undertaken by enterprises; it necessitates the establishment of robust collaborative partnerships with universities and scientific research institutions. Enterprises must actively pursue deep collaborations with these entities, collectively undertaking fundamental and applied research and development endeavors in big data technology. Through this collaborative paradigm, enterprises can promptly access cutting-edge technological advancements and research trends while leveraging the intellectual capital of academia to bolster their own research capabilities and innovation prowess. Such collaborative efforts are integral to continually exploring and innovating application scenarios and solutions for big data technology within accounting and auditing realms, ultimately driving sustainable growth and competitiveness.

### 4.2. The improvement of accounting standards and systems

In the backdrop of the big data era, the government and relevant institutions shoulder crucial oversight and guidance responsibilities. The rapid proliferation and extensive integration of big data technology have catalyzed profound transformations across various industries, particularly within the realms of accounting and finance. Hence, these institutions must maintain acute sensitivity and foresight, closely monitoring the evolving landscape and latest trends within the big data era. To ensure that accounting standards and systems remain abreast of contemporary advancements, it is imperative for the government and relevant bodies to periodically assess the relevance of existing standards and systems. Timely revisions and enhancements should be made in accordance with the distinct characteristics of big data technology and the evolving demands of industries. This iterative process necessitates a comprehensive understanding of how big data impacts the quality of accounting information, processing methodologies, and disclosure requirements. It is imperative to ensure that these standards and systems effectively guide practical application and safeguard the interests of all stakeholders. Enterprises, as the primary executors of accounting standards and systems, play a pivotal role in this process. Their active engagement in the formulation phase is paramount. Drawing upon their practical experiences in leveraging big data, enterprises can offer valuable insights and recommendations to governmental and institutional entities. Such contributions serve to enhance the practicality and feasibility of standards and systems, fostering improved communication and collaboration between enterprises and the government. This collaborative effort is integral to fostering the healthy evolution of the accounting industry in tandem with the demands of the big data era.

### 4.3. The strengthening of information security and privacy protection measures

In the digital age, enterprises must establish a robust and efficient information security management system to safeguard their invaluable data assets and mitigate potential threats. At the heart of this system lies the reinforcement of security measures such as data encryption and access control, aimed at preserving the confidentiality, integrity, and availability of enterprise data. Data encryption technology plays a pivotal role in safeguarding sensitive information both during transmission and storage, even amidst potential interception risks. Meanwhile, access control technology ensures that only authorized users can access designated data resources through stringent identity authentication and authority management protocols, effectively mitigating internal breaches and unauthorized intrusions. However, while technical safeguards are crucial, human factors cannot be overlooked. Enterprises must prioritize information security awareness training for employees, fostering a deep understanding of the significance of information security and privacy protection. Through regular training and education initiatives, employees can acquire the skills to identify and address various information security risks, adhere to information security policies in their daily tasks, and collaboratively fortify the enterprise's information security defenses [3].

### 4.4. The improvement of auditors' quality and ability

In today's dynamic business landscape, the role of auditors transcends traditional account verification and risk assessment. Enterprises are increasingly recognizing the crucial need for comprehensive and nuanced training to elevate auditors' professional caliber and breadth of expertise. Such training must extend beyond conventional audit practices to encompass emerging skills like big data analysis, data mining, and the utilization of cutting-edge audit tools. In the age of big data, proficiency in data analysis technology stands as the cornerstone of auditor competency. Consequently, enterprises must prioritize providing auditors with specialized training and hands-on experience to adeptly leverage diverse data analysis tools. This proficiency empowers auditors to effectively distill invaluable insights from vast datasets, thereby furnishing robust support for strategic decision-making within enterprises. Equally paramount is the incorporation of data mining methodologies into auditor training initiatives. Proficiency in these techniques equips auditors with the acumen to discern latent risks and issues with heightened precision, thereby augmenting the efficiency and accuracy of audit processes. Moreover, fostering auditors' communication aptitude and collaborative ethos assumes critical significance. Given the interdisciplinary nature of auditing, adept communication skills facilitate seamless information dissemination and prompt issue resolution. Likewise, cultivating a spirit of teamwork among auditors fosters cohesion and enhances the efficacy of audit teams, as aligned goals and values propel collective efforts towards optimal outcomes.

## 5. Conclusion

The advent of the big data era has profoundly reshaped the landscape of accounting and auditing, presenting a plethora of opportunities alongside significant challenges. In order to navigate this transformative period effectively, it is imperative for enterprises and governments to proactively address these challenges. Leveraging the inherent advantages of big data technology is paramount to driving forward the evolution and innovation within the realms of accounting and auditing. To collectively foster the healthy progression of accounting and auditing within the big data era, it is essential to bolster efforts in technology research and application. Additionally, enhancing accounting standards and systems, fortifying information security and privacy measures, and augmenting the proficiency and caliber of auditors are pivotal steps. Through concerted actions in these areas, stakeholders can collaboratively steer the trajectory of accounting and auditing towards sustainable growth and excellence in the dynamic landscape of big data.